 \documentstyle[12pt]{article}
\let\la=\label  
  
 \def\bd{\begin{document}} \def\ed{\end{document}}
\def\ds{\documentstyle} \let\fr=\frac \let\bl=\bigl \let\br=\bigr
\let\Br=\Bigr \let\Bl=\Bigl
\let\bm=\bibitem
\let\na=\nabla
\let\pa=\partial \let\ov=\overline
\newcommand{\be}{\begin{equation}}
\newcommand{\ee}{\end{equation}}
\def\ba{\begin{array}}
\def\ea{\end{array}}
\newcommand{\ho}[1]{$\, ^{#1}$}
\newcommand{\hoch}[1]{$\, ^{#1}$}
\newcommand{\bea}{\begin{eqnarray}}
\newcommand{\eea}{\end{eqnarray}}
\newcommand{\ra}{\rightarrow}
\newcommand{\lra}{\longrightarrow}
\newcommand{\Lra}{\Leftrightarrow}
\newcommand{\ap}{\alpha^\prime}
\newcommand{\bp}{\tilde \beta^\prime}
\newcommand{\tr}{{\rm tr} }
\newcommand{\Tr}{{\rm Tr} }
\newcommand{\NP}{Nucl. Phys. }
\newcommand{\tamphys}{\it  Center
for Theoretical Physics, Texas A\&M University, College Station, Texas 77843.
}

\newcommand{\auth}{M. J. Duff }

\begin{document}

\hfill{CTP-TAMU-32/95}

\hfill{hep-th/9509106}

\vspace{24pt}

\begin{center}
{ \large {\bf ELECTRIC/MAGNETIC DUALITY AND ITS STRINGY ORIGINS\footnote
{Based on review
talks delivered at the PASCOS 95 conference, Johns Hopkins University, 
Baltimore, March
1995 and the SUSY 95 conference, Ecole Polytechnique, Paris, June 1995.  
Research
supported in part by NSF Grant  PHY-9411543.} }}

\vspace{36pt}

\auth

\vspace{10pt}

{\tamphys}

\vspace{44pt}

\underline{ABSTRACT}

\end{center}

We review electric/magnetic duality in $N=4$ (and certain $N=2$) globally
supersymmetric gauge theories and show how this duality, which relates 
strong to weak coupling, follows as a consequence of a string/string duality. 
Black holes, eleven dimensions and supermembranes also have a part to play 
in the big picture.

{\vfill\leftline{}\vfill
\leftline
\pagebreak

\section{Introduction}
\la{Introduction}

Two of the hottest topics in theoretical high-energy physics at the moment are:

\noindent 1) {\it Electric/magnetic duality} in $D=4$ dimensional globally
supersymmetric gauge theories, whereby the long distance behavior of strongly
coupled {\it electric} theories are described in terms of weakly coupled
{\it magnetic} theories. This sheds new light on quark confinement, the Higgs
mechanism and even pure mathematics.

\noindent 2) {\it String/string duality} in $D\leq6$ dimensional superstring
theory, whereby the same physics is described by two apparently different
theories: one a heterotic string and the other a Type $II$ string. 

\noindent Here I wish to review both and show that the former follows as a
consequence of the latter.  We shall also discover that black holes, $D=11$
dimensions and extended objects with more than one spatial dimension (the
{\it super p-branes}) also make their appearance.

In discussing duality in gauge theories it is important to distinguish between
{\it exact duality} and {\it effective duality}.  We shall take the phrase
{\it exact duality} to refer to the conjectured $SL(2,Z)$ symmetry that acts on
the gauge coupling constant $e$ and theta angle $\theta$:
\be
S \rightarrow \frac{aS+b}{cS+d}
\la{sl2zs}
\ee
where $a,b,c,d$ are integers satisfying $ad-bc=1$ and where
\be
S=S_1+iS_2=\frac{\theta}{2\pi} + i\frac{4\pi}{e^2}
\la{S}
\ee
This is also called electric/magnetic duality because the integers $m$ and
$n$ which characterize the magnetic charges $Q_m=n/e$ and electric charges
$Q_e=e(m+n\theta/2\pi)$ of the particle spectrum transform as
\be
\left( \begin{array}{c}
m\\
n
\end{array}
\right)
\rightarrow
\left( \begin{array}{cc}
a&b\\
c&d
\end{array}
\right)
\left( \begin{array}{c}
m\\
n
\end{array}
\right)
\la{charges}
\ee
Such a symmetry would be inherently non-perturbative since, for $\theta=0$
and with $a=d=0$ and $b=-c=-1$, it reduces to the strong/weak coupling duality  
\[
{e}^2/4\pi \rightarrow4\pi/{e}^2
\]
\be
n\rightarrow m, m\rightarrow -n
\label{simple}
\ee
This in turn means that the coupling constant cannot get renormalized in
perturbation theory and hence that the renormalization group
$\beta$-function vanishes
\be
\beta(e)=0
\ee
This is guaranteed in $N=4$ supersymmetric Yang-Mills and also happens in
certain $N=2$ theories.  Thus the duality idea is that the theory
may equivalently be described in one of two ways. In the conventional way, the
W-bosons, Higgs bosons and their fermionic partners are the
electrically charged elementary particles and the magnetic monopoles emerge
as soliton solutions of the field equations.  In the dual description, however,
it is the monopoles which are elementary and the electrically charged particles
which emerge as the solitons. Historically, the exact duality (\ref{simple}) was
first conjectured by Montonen and Olive \cite{Goddard,Montonen,Olive,Osborn} and
then generalized to include the theta angle in 
\cite{Cardy1,Cardy2,Shapere,Sen1,Girardello,Vafawitten1,Seibergwitten1,%
Seibergwitten2}. 
The $SL(2,Z)$ then means that there are, in fact, infinitely many equivalent 
descriptions.

We shall take the phrase {\it effective duality} to refer to the more
realistic $N=1$ and $N=2$ gauge theories, for which $\beta(e)\neq 0$ and which
exhibit no exact $SL(2,Z)$ symmetry, but for which there exists two
different versions, with different gauge groups and quark representations,
leading to the same physics. The weak coupling region of one theory is mapped
into the strong coupling region of the other.  Once again there is a soliton
interpretation.  This {\it effective duality} has been pioneered by Seiberg and
Witten \cite{Seibergwitten1,Seibergwitten2,Seibergpower,Seibergelectric} and it
shows great promise for the understanding of quark confinement and chiral
symmetry breaking. Both arise from the condensation of the magnetic monoples.
Moreover, it is these effectively dual theories which have proved most 
valuable in
proving new results in the topology of four-manifolds \cite{Wittenmonopole}.  
However, my
main purpose in this review is to explain how electric-magnetic duality 
follows by
embedding these Yang-Mills theories in a superstring theory and to date
most progress in this direction has been made in the context of exactly dual
theories. In what follows, therefore I will focus on exact duality and 
confine the
discussion of the stringy origins of effective duality to a few remarks at the
end.  This will also render possible the otherwise impossible task of reviewing
the two hottest topics in just a few pages.

\section{Monopoles of $N=4$ and $N=2$ Yang-Mills}
\la{Monopoles}

Consider the action describing the bosonic sector of the unique $N=4$
supersymmetric $SU(2)$ Yang-Mills theory:
\be
S=\int d^4x\left(-\frac{1}{4e^2}TrF_{\mu\nu}F^{\mu\nu}-\frac{1}{2}TrD_{\mu}\Phi
D^{\mu}\Phi- V(\Phi) +\frac{\theta}{16\pi^2}TrF_{\mu\nu}{\tilde
F}^{\mu\nu}\right) 
\ee
The potential $V(\Phi)$ admits spontaneous symmetry breaking VEVs for the
scalar fields $<Tr\Phi^2>=v^2\neq 0$ which break the $SU(2)$ down to $U(1)$.  
This
means that the theory admits BPS monopoles. Both the elementary states and the
monopoles belong to short $16$-dimensional $N=4$ supermultiplets and their
masses saturate a Bogomol'nyi bound:
\be
M^2=(4\pi)^2(m,n)\frac{v^2}{S_2} 
\left( \begin{array}{cc}
1&S_1\\
S_1&|S|^2 
\end{array}
\right)
\left( \begin{array}{c}
m\\
n
\end{array}
\right)
\la{mass}
\ee
This universal mass formula (\ref{mass}) is manifestly invariant under the
$SL(2,Z)$ transformations (\ref{sl2zs}) and (\ref{charges}).  Although it was
obtained by semiclassical reasoning, $N=4$ non-renormalization theorems
ensure that it is exact in the full quantum theory. The duality conjecture is
that this $SL(2,Z)$ is not only an exact symmetry of mass spectrum but of the
entire quantum theory.

Strong evidence for this conjecture was provided by Sen \cite{Sen1} who
pointed out that given a purely electrically charged state $(m=1,n=0)$, 
$SL(2,Z)$
implies the existence of a state $(p,q)$ with $p$ and $q$ relatively prime
integers (i.e having no common divisor). Sen then went on to construct
explicitly a dyonic solution with charges $(1,2)$ in complete agreement with
the conjecture.  Further evidence was supplied by Vafa and Witten
\cite{Vafawitten1}, by studying partition functions of the twisted 
$N=4$ theory on
various $4$-manifolds, and also by Girardello et al \cite {Girardello}.

Originally, it was thought that this exact duality between the elementary
particles and the monopoles would work only for $N=4$, where both belong to
the unique short supermultiplet and was shown explicitly not to work for
the pure $N=2$ theory \cite{Osborn}.  However, the monopole spectrum is 
consistent
with exact duality in the $\beta(e)=0$ case of $N=2$ Yang-Mills with gauge group
$SU(2)$ coupled to four hypermultiplets in the fundamental representation, 
as was
shown in \cite{Seibergwitten1,Seibergwitten2} for the one-monopole sector 
($q=1$)
and recently in \cite{Sethi,Cederwall,Gauntlett2} for the two-monopole sector
($q=2$).  There are more subtleties in the $N=2$ case.  For example, when 
the $8$
real fermions in the hypermultiplets transform under $SL(2,Z)$, they must also
transform under an $SO(8)$ triality.  

\section{$S$-duality in string theory}
\la{Sdual}

In string theory the roles of the theta angle
$\theta$ and coupling constant $e$ are played by the VEVs of the the
four-dimensional axion field $a$ and dilaton field $\eta$: 
\be
<a>=\frac{\theta}{2\pi}
\ee
\be
{e}^2/4\pi=<e^{\eta}>=8G/\alpha'
\ee
Here $G$ is Newton's constant and $2\pi\alpha'$ is the inverse string tension.
Hence $S$-duality (\ref{sl2zs}) now becomes a transformation law for the
axion/dilaton field $S$: 
\be
S=S_1+iS_2={\rm a}+ie^{-\eta}
\la{Sfield}
\ee

The $S$-duality conjecture in string theory has its origins in supergravity. In
the late 70s and early 80s, it was realized that compactified supergravity
theories exhibit non-compact global symmetries \cite{Scherk,Cremmer,Marcus,%
Dufffradkin} e.g
$SL(2,R)$, $O(22,6)$, $O(24,8)$, $E_7$, $E_8$, $E_9$, $E_{10}$.  In 1990 it was
conjectured \cite{Luduality,Dufflu} that discrete subgroups of all these
symmetries   should be promoted to duality symmetries of either heterotic or 
Type
$II$ superstrings. The case for $O(22,6;Z)$ had already been made. 
This is the well-established target space duality, sometimes called {\it
$T$-duality} \cite{Giveonreview}.  Stronger evidence for a strong/weak coupling
$SL(2,Z)$ duality in string theory was subsequently provided in
\cite{Font,Rey,Kalara,Sen2,Sen3,Schwarz1,Schwarz2,Binetruy,Khurifour,Sen4,%
Rahmfeld,Gauntlett1,Khuristring},
stronger evidence for the combination of $S$ and $T$-duality into an $O(24,8;Z)$
in heterotic strings was provided in \cite{Rahmfeld,Duffclassical,Sen5,%
Rahmfeld2} and
stronger evidence for their combination into a discrete $E_7$ in Type $II$ 
strings
was provided in \cite{Hulltownsend}, where it was dubbed {\it $U$-duality}.  

Let us first consider $T$-duality and focus just on the moduli fields that arise
in  compactification on a $2$-torus of a $D=6$ string with dilaton $\Phi$, 
metric
$G_{MN}$ and $2$-form potential $B_{MN}$ with $3$-form field strength $H_{MNP}$.
Here the $T$-duality is just $O(2,2;Z)$.  Let us parametrize the compactified
($m,n=4,5$) components of string metric and 2-form as   
\be
G_{mn}=e^{\rho-\sigma}\left( \begin{array}{cc}
e^{-2\rho}+c^2&-c\\
-c&1
\end{array}\right)
\ee
and
\be
B_{mn}=b\epsilon_{mn}
\ee
The four-dimensional shifted dilaton $\eta$ is given by   
\be
e^{-\eta}=e^{-\Phi}\sqrt{det G_{mn}}=e^{-\Phi -\sigma}
\ee
and the axion field ${\rm a}$ is defined by 
\be
\epsilon^{\mu\nu\rho\sigma}\partial_{\sigma}{\rm a}=
\sqrt{-g}e^{-\eta}g^{\mu\sigma}g^{\nu\lambda}g^{\rho\tau}H_{\sigma\lambda\tau}
\ee
where $g_{\mu\nu}=G_{\mu\nu}$ and $\mu,\nu=0,1,2,3$. We further define the 
complex
Kahler form field $T$ and the complex structure field $U$ by  
\[
T=T_1+iT_2=b+ie^{-\sigma}
\]
\be
U=U_1+iU_2=c+ie^{-\rho}
\ee
Thus this $T$-duality may be written as
\be
O(2,2;Z)_{TU} \sim SL(2,Z)_T \times SL(2,Z)_U
\la{Tsplit}
\ee
where $SL(2,Z)_T$ acts on the $T$-field and $SL(2,Z)_U$ acts on the $U$-field
in the same way that $SL(2,Z)_S$ acts on the $S$-field in (\ref{sl2zs}).  In
contrast to $SL(2,Z)_S$, $SL(2,Z)_T \times SL(2,Z)_U$ is known to be not 
merely a
symmetry of the supergravity theory but an exact string symmetry order by order
in string perturbation theory. $SL(2,Z)_T$ does, however, contain a
minimum/maximum length duality mathematically similar to (\ref{simple}) 
\be
R \rightarrow \alpha'/R
\la{R}
\ee
where $R$ is the compactification scale given by
\be
\alpha'/R^2=<e^{\sigma}>.
\ee

Even before compactification, the Type $IIB$ supergravity exhibits an $SL(2,R)$
whose  discrete subgroup has been conjectured to be a non-perturbative
symmetry of the Type $IIB$ string \cite{Callan2,Hulltownsend}.  We shall
refer to this duality as $SL(2,Z)_X$ to distinguish it from the others.  
Combining
this with the known $T$-duality of the four dimensional theory obtained by
compactification on $T^6$ leads to the $E_7$.  So the explanation for 
$U$-duality
devolves upon the explanation for this $SL(2,Z)_X$.  We shall return to this in
section (\ref{triality}).

\section{$S$-duality from $D=6$ string/string duality}
\la{string}

Let us now investigate how both $N=4$ and $N=2$ exact electric/magnetic duality
follows from string theory.  As discussed above, there is a formal similarity
between this symmetry and that of $T$-duality.  It was argued in
\cite{Duffstrong} that these mathematical similarities between $SL(2,Z)_S$ and
$SL(2,Z)_T$ are not coincidental. Evidence was presented in favor of the idea
that the physics of the fundamental string in six spacetime dimensions may
equally well be described by a dual string and that one emerges as a soliton
solution of the other \cite{Luloop,Lublack,Minasian1,Duffstrong,Senssd,Harvey}.
The string equations admits the singular {\it elementary} string solution
\cite{Dabholkar}  
\[
ds^2= (1-k^2/r^2)[-d\tau^2+d\sigma^2 + (1-k^2/r^2)^{-2}dr^2
+r^2d\Omega_{3}{}^2]
\]
\[
e^{\Phi}=1-k^2/r^2
\]
\be
e^{-\Phi}*H_3=2k^2\epsilon_3
\la{fund}
\ee
where
\be
k^2=\kappa^2 T/\Omega_3
\ee
$T=1/2\pi \alpha'$ is the string tension, $\Omega_3$ is the volume of $S^3$ and
$\epsilon_3$ is the volume form. It describes an
infinitely long string whose worldsheet lies in the plane $X^0=\tau,X^1
=\sigma$.  Its mass per unit length is given by 
\be
 M= T<e^{\Phi/2}>
\ee
and is thus heavier for stronger string coupling, as one would expect for
a fundamental string.  The string equations also admit the non-singular
{\it solitonic} string solution
\cite{Lublack,Minasian1}
\[
ds^2= -d\tau^2+d\sigma^2 + (1-\tilde k^2/r^2)^{-2}dr^2 + r^2d\Omega_{3}{}^2
\]
\[
e^{-\Phi}=1-\tilde k^2/r^2
\]
\be
H_3=2\tilde k^2\epsilon_3
\la{sol}
\ee
whose tension $\tilde T=1/2\pi {\tilde \alpha}'$ is given by
\be
\tilde k^2=\kappa^2 \tilde T/\Omega_3
\ee
Its mass per unit length is given by
\be
\tilde {M}= \tilde T <e^{-\Phi/2}>
\ee
and is thus heavier for weaker string coupling, as one would expect
for a solitonic string. Thus we see that the solitonic string differs from the
fundamental string by the replacements 
\[
\Phi \rightarrow \tilde \Phi=-\Phi
\]
\[
G_{MN} \rightarrow \tilde G_{MN}=e^{-\Phi}G_{MN}
\] 
\[
H \rightarrow \tilde H=e^{-\Phi}*H
\]
\be
\alpha' \rightarrow \tilde \alpha'
\la{dual}
\ee
The Dirac quantization rule $eg=2\pi n$ ($n$=integer) relating the Noether
``electric'' charge
\be
e=\frac{1}{\sqrt{2}\kappa}\int_{S^3}e^{-\Phi}*H_3
\ee
to the topological ``magnetic'' charge
\be
g=\frac{1}{\sqrt{2}\kappa}\int_{S^3}H_3
\ee
translates into a quantization condition on the two tensions:
\be
2\kappa{}^2=n(2\pi)^3\alpha'\tilde \alpha'\,\,\,\,\,\,n=integer
\la{Dirac1}
\ee
where $\kappa$ is the six-dimensional gravitational constant. 
Both the string and dual string soliton solutions break half the
supersymmetries, both saturate
a Bogomol'nyi bound between the mass and the charge. These
solutions are the extreme mass equals
charge limit of more general two-parameter black string solutions
\cite{Horowitz1,Lublack}.

We now make the major assumption of string/string duality: the dual
string may be regarded as a fundamental string in its own right with a 
worldsheet
action that couples to the dual variables and has the dual tension
given in (\ref{dual}). It follows that the dual string equations admit the dual
string (\ref{sol}) as the fundamental solution and the
fundamental string (\ref{fund}) as the dual solution.	When expressed 
in terms of
the dual metric, however, the former is singular and the latter non-singular. It
also follows from (\ref{Dirac1}) that in going from the fundamental string to
the dual string and interchanging  $\alpha'$ with 
${\tilde\alpha}'=2\kappa^2/(2\pi)^3\alpha'$, one also interchanges the roles of
worldsheet and spacetime loop expansions. Moreover, since the dilaton enters the
dual string equations with the opposite sign to the fundamental string, it was
argued in \cite{Luloop,Lublack,Minasian1} that in $D=6$ the strong coupling 
regime
of the string should correspond to the weak coupling regime of the dual string:
\be
{\rm g}_6{}^2/(2\pi)^3 = <e^{\Phi}>=(2\pi)^3/{\tilde{\rm g}_6}^2
\la{coupling}
\ee
where ${\rm g_6}$ and $\tilde{\rm g}_6$ are the six-dimensional string
and dual string loop expansion parameters.

On compactification to four spacetime dimensions, the two theories
appear very similar, each acquiring an $O(2,2;Z)$ target space duality.	One's
first guess might be to assume that the strongly coupled
four-dimensional fundamental
string corresponds to the weakly coupled dual string, but in fact
something more subtle and
interesting happens: the roles of the $S$ and $T$ fields are
interchanged \cite{Khurifour} so that the
strong/weak coupling $SL(2,Z)_S$ of the fundamental string emerges as a
subgroup of the target space
duality of the dual string:
\be
O(2,2;Z)_{SU} \sim SL(2,Z)_S \times SL(2,Z)_U
\la{Ssplit}
\ee
This {\it duality of dualities} is summarized in Table (1).
\begin{table}
$
\begin{array}{lll}
&Fundamental \, string&Dual \, string\\
&&\\
T-duality&O(2,2;Z)_{TU} &O(2,2;Z)_{SU}\\
&\sim SL(2,Z)_T \times	SL(2,Z)_U&\sim SL(2,Z)_S\times	SL(2,Z)_U\\
Moduli&T={\rm b}+ie^{-\sigma}&S={\rm a}+ie^{-\eta}\\
&{\rm b}=B_{45}&{\rm a}=\tilde B_{45}\\
&e^{-\sigma}=\sqrt{detG_{mn}}&e^{-\eta}=\sqrt{det \tilde{G}_{mn}}\\
Worldsheet \, coupling&<e^{\sigma}>=\alpha'/R^2&<e^{\eta}>={\rm g}^2/2\pi\\
Large/small \, radius &R\rightarrow \alpha'/R&{\rm g}^2/2\pi\rightarrow
2\pi/{\rm g}^2\\
S-duality&SL(2,Z)_S&SL(2,Z)_T\\
Axion/dilaton&S={\rm a}+ie^{-\eta}&T={\rm b}+ie^{-\sigma}\\
&d{\rm a}=e^{-\eta}*H&d{\rm b}=e^{-\sigma}\tilde{*} \tilde{H}\\
&e^{-\eta}=e^{-\Phi}\sqrt{detG_{mn}}&e^{-\sigma}=e^{\Phi}\sqrt{det
\tilde{G}_{mn}}\\
Spacetime \, coupling&<e^{\eta}>={\rm g}^2/2\pi&<e^{\sigma}>=\alpha'/R^2\\
Strong/weak \, coupling&{\rm g}^2/2\pi\rightarrow 2\pi/{\rm g}^2&R
\rightarrow \alpha'/R
\end{array}
$
\label{table}
\caption{Duality of dualities}
\end{table}
As a consistency check, we note that since $(2\pi R)^2/2\kappa^2=1/16\pi G$
the Dirac
quantization rule (\ref{Dirac1}) becomes (choosing $n$=1)
\be
8GR^2=\alpha'\tilde \alpha'
\la{Dirac2}
\ee
Invariance of this rule now requires that a strong/weak coupling
transformation on the fundamental
string ($8G/\alpha'\rightarrow \alpha'/8G$) must be accompanied by a
minimum/maximum length
transformation of the dual string ($\tilde \alpha'/R^2 \rightarrow
R^2/\tilde \alpha'$), and vice
versa.

\section{Electric/magnetic duality and extreme black holes}
\la{electric}

Turning now to the electric and magnetic fields, these fall naturally into two
categories: (1) the gauge fields already present in the $D=6$ string theory and
whose details will depend on how we arrived at this theory; (2) the
$U(1)^4$ fields which arise in going from $6$ to $4$ dimensions on a generic
$T^2$ and which appear in $G_{\mu n}$ ( the Kaluza-Klein gauge fields) and 
$B_{\mu
n}$ (the winding gauge fields). We begin with (2)
which are easier to discuss.	 The target space $T$-duality and $U$-duality
and the strong-weak coupling $S$-duality transform the four field strengths
${{F}}_{\mu\nu}$ and their duals ${\tilde {F}}_{\mu\nu}$ as 
\be
{F}_{\mu\nu}\rightarrow (\omega_T{}^{-1} \times  \omega_U{}^{-1}) {F}_{\mu\nu}
\ee
\be
\left(
\begin{array}{c}
{{F}}_{\mu\nu}\\
{\tilde {F}}_{\mu\nu}
\end{array}
\right)
\rightarrow	\omega_S^{-1}
\left(
\begin{array}{c}
{{F}}_{\mu\nu}\\
{\tilde {F}}_{\mu\nu}
\end{array}
\right)
\ee
where $\omega_S$, $\omega_T$ and $\omega_U$ are the respective $SL(2,Z)$
matrices. Thus $T$-duality transforms Kaluza-Klein electric charges
$({F}^3,{F}^4)$ into winding electric charges $({F}^1,{F}^2)$ (and
Kaluza-Klein magnetic charges into winding magnetic charges), $U$-duality
transforms the Kaluza-Klein and winding electric charge of one circle
$({F}^3,{F}^2)$ into those of the other $({F}^4,{F}^1)$ (and similarly for
the magnetic charges) but $S$-duality transforms Kaluza-Klein electric charge
$({F}^1,{F}^2)$ into winding magnetic charge $({\tilde {F}}_3,{\tilde
{F}}_4)$ (and winding electric charge into Kaluza-Klein magnetic charge).  In a
way which should now be obvious, an entirely similar story applies to the dual
theory with $T$ and $S$ exchanging roles. Note that the solitonic magnetic {\it
H-monopoles} \cite{Khuri,Liu} of the fundamental string are the fundamental
electric winding states of the dual string \cite{Khurinew,Rahmfeld}. The
Kaluza-Klein states are common to both.

It is here that the black hole connection enters.  In \cite{Khurinew} it was
shown that these $H$-monopoles are in fact magnetically charged black holes
in the limiting case where the charge equals the mass. By $T$-duality, they are
related to the Kaluza-Klein monopoles which are also extreme (mass=magnetic
charge) black holes.  But by $S$-duality if these solitonic magnetically charged
string states are black holes, then the elementary electrically charged string
states must also be extreme (mass=electric charge) black holes \cite{Rahmfeld}
because they belong to the same $SL(2,Z)$ doublet!  To be precise, if we 
denote by
$N_L$ and $N_R$ the number of left and right oscillators of the heterotic 
string,
then the extreme black holes correspond to $N_R=1/2$. In the literature, black
holes are frequently labeled by a parameter $a$ which describes how the Maxwell
field couples to a single scalar field formed from a combination of dilaton and
moduli. The $(N_R=1/2,N_L=1)$ states yield $a=\sqrt{3}$ and the 
$(N_R=1/2,N_L>1)$ states
(with vanishing left-moving internal momentum) yield $a=1$.  
One might worry that this
identification of Bogomoln'yi string states with extreme black holes 
works only at the
level of the charge and mass spectrum, so it is worth emphasing that it 
extends also to
the gyromagnetic ratios \cite{Rahmfeld,Senblack}. Moreover, further 
{\it dynamical}
evidence  has been supplied in \cite{Myers}, where it was shown 
(in the limit of low
velocities) that the $a=\sqrt{3}$ and $a=1$ extreme black holes have the 
same scattering
amplitudes as the $(N_R=1/2,N_L=1)$ and  $(N_R=1/2,N_L>1)$ string states. 
It has recently
been shown that there are also {\it massless} charged black holes
corresponding to $(N_R=1/2,N_L=0)$ states
\cite{Behrndt,Linde,Cvetic}.  

\section{Symmetry enhancement}
\la{Symmetry}

All of our discussions of the compactifying torus $T^2$ have so far
assumed that we are at a generic point in the moduli space of vacuum
configurations and that the unbroken gauge symmetry in going from $D=6$ to $D=4$
is the abelian $U(1)^4$.	However, we know that at special points in 
moduli space
two of the four $U(1)$s may be enhanced \cite{Green} to a simply laced rank $2$
non-abelian group. For $T=U$; $T=U=i$; $T=U=exp({2\pi i/3})$, the enhanced
symmetries are $SU(2) \times U(1)$, $SU(2) \times SU(2)$ and
$SU(3)$, respectively. String/string duality now suggests a
new (non-perturbative) phenomenon, however.	In theories with an
$S \leftrightarrow T$ symmetry, such as the one discussed in section 
(\ref{pairs}) 
below, a similar enhancement of the dual gauge symmetry also occurs in the 
dual theory when
$S=U$ and $S=T$ \cite{Duffstrong,Rahmfeld3}. 

In order to discuss the gauge fields present already in $D=6$ and the questions
of symmetry enhancement there, it is first necessary to be more specific about
the nature of the dual string.  

\section{A concrete $N=4$ example}
\la{concrete}

When we say that one string is dual another, exactly which strings are we
talking about?  After all, {\it any} string in $D=6$ will exhibit the 
fundamental
and solitonic string solutions of section (\ref{string}) because all strings
couple to the metric, $2$-form and dilaton.  The solitonic zero-modes that
describe the field content of the dual string worldsheet will, however, depend
crucially on the nature of the fundamental string.  In the author's opinion,
one of the most important unsolved problems in string/string duality is that 
there
is as yet no well-defined criterion for deciding when a solitonic string
should be promoted to the status of being fundamental in its own right. Must it
be critical, for example? At the moment, therefore, the game is to play safe and
focus only on those dual solitonic strings that correspond to fundamental
critical string theories that we already know.  The example that has 
attracted the
most interest is the conjecture put forward by Hull and Townsend
\cite{Hulltownsend} and by Witten \cite{Wittenvarious} which states that the 
$D=10$
heterotic string compactified to $D=6$ on $T^4$  is dual to the $D=10$ 
Type $IIA$
string compactified to $D=6$ on $K3$.  $K3$	is	a 
four-dimensional	compact closed
simply-connected manifold.	It	is equipped	with	
a self-dual	metric	and	hence its
holonomy	group	is $SU(2)$.	It	was first	
invoked	in a	Kaluza-Klein	context	in
\cite{Nilsson2,Townsendk3} where	it	was used as	a	
way	of compactifying	$D=11$
supergravity to	$D=7$ and $D=10$ supergravity to	$D=6$.	
Half the spacetime
supersymmetry remains	unbroken	as	a consequence	of the	
$SU(2)$ holonomy,	and	hence
the Type $IIA$ theory	gives rise	to an	$N=2$	string	in $D=6$. 
In fact, in 1986, it
was pointed out \cite{Nilsson1} that $D=11$ supergravity compactified on 
$K3\times T^{n-3}$  \cite{Nilsson2} and the $D=10$ heterotic string compactified
on $T^{n}$ \cite{Narain} have the same moduli spaces of vacua, namely  
\begin{equation} {\cal M}=\frac{SO(16+n,n)}{SO(16+n) \times SO(n)}
\label{moduli}
\end{equation}
It was subsequently confirmed \cite{Seiberg,Aspinwall1}, in the context 
of the $D=10$
Type $IIA$ theory compactified on $K3\times T^{n-4}$, that this 
equivalence holds
globally as well as locally.  These ``coincidences" lend further credence to the
conjecture.  We shall have more to say on the role of $D=11$ supergravity in
section (\ref{web}).

Further $T^2$
compactification yields a dual pair of $D=4,N=4$ strings that are related by the
interchange of $S$ and $T$ discussed in section \ref{string}.  In 
particular, the
$S$-duality of the Type $IIA$ string follows automatically from the 
$T$-duality of
the heterotic string.  Moreover, just as special points in moduli space lead to
enhanced gauge symmetries on the heterotic side, string/string duality 
implies that
symmetry enhancement must occur at special $K3$ points on the $IIA$ side
\cite{Hulltownsend,Wittenvarious,Aspinwall3}. If we now take the global 
limit of this
theory, we find  $N=4$ Yang-Mills theories with the desired $SL(2,Z)$.  
Of course, this
global limit involves starting with a theory of gravity and then switching 
the gravity
off.  Witten \cite{Wittencomments} has suggested that a more direct way of 
embedding the
Yang-Mills theory in string theory in order to derive the $SL(2,Z)$ would be 
via the
{\it anti-self-dual string} \cite{Rahmfeld2} in the sense that this may be 
the minimal
manifestly $S$-dual extension of the $N=4$ super Yang-Mills theory. 

\section{Four-dimensional string/string/string triality}
\la{triality}

We have seen that in six spacetime dimensions, the heterotic string is dual to a
Type $IIA$ string.  On further toroidal compactification to four spacetime
dimensions, the heterotic string acquires an $SL(2,Z)_S$
strong/weak coupling duality and an $SL(2,Z)_T \times
SL(2,Z)_U$ target space duality acting on the dilaton/axion,
complex Kahler form and the complex structure fields $S,T,U$
respectively.  Strong/weak duality in $D=6$ interchanges the roles of
$S$ and $T$ in $D=4$ yielding a Type $IIA$ string with fields $T,S,U$.

However, as discussed in section (\ref{Sdual}), the target space symmetry of the
heterotic theory also contains an $SL(2,Z)_U$ that acts on $U$, the complex
structure of the torus.  This suggests that, in addition to these $S$ and $T$
strings there ought to be a third {\it $U$-string} whose axion/dilaton field 
is $U$
and whose strong/weak coupling duality is $SL(2,Z)_U$.  From a
$D=6$ perspective, this seems strange since, instead of (\ref{dual}),
we now interchange $G_{45}$ and $B_{45}$. Moreover, of the two electric
field strengths which become magnetic, one is a winding gauge field and
the other is Kaluza-Klein! So such a duality has no $D=6$ Lorentz
invariant meaning.  In fact, this $U$ string is a Type $IIB$ string, a
result which may also be understood from the point of view of
mirror symmetry:  interchanging the roles of Kahler form and complex structure
(which is equivalent to inverting the radius of one of the two circles)
is a symmetry of the heterotic string but takes Type $IIA$ into Type
$IIB$ \cite{Dai,Dine}.  In summary, if we denote the heterotic, $IIA$
and $IIB$ strings by $H,A,B$ respectively and the axion/dilaton,
complex Kahler form and complex structure by the triple $XYZ$ then we
have a triality between the $S$-string ($H_{STU}=H_{SUT}$), the
$T$-string ($B_{TUS}=A_{TSU}$) and the $U$-string ($A_{UST}=B_{UTS}$)
\cite{Rahmfeld3}. Related results have been obtained independently in
\cite{Aspinwall2} and \cite{Kaloper}. Each string in $D=4$ will then 
exhibit the same total symmetry 
\be
SL(2,Z)_S \times O(6,22;Z)_{TU} \supset
SL(2,Z)_S \times SL(2,Z)_T \times SL(2,Z)_U
\ee
with the $28$ gauge field strengths and their duals transforming as a $(2,28)$. 
Of course, there will be different interpretations for the three $SL(2,Z)$
factors.  So although there is a discrete symmetry under $T \leftrightarrow U$
interchange, there is no such $U \leftrightarrow S$ or $S \leftrightarrow T$
symmetry.  As discussed in \cite{Duffstrong}, it is the degrees of freedom
associated with going from $10$ to $6$ which are responsible for this lack of
$S$--$T$--$U$ democracy.  This will also be reflected in the
Bogomol'nyi spectrum of electric and magnetic states that belong to the
short and intermediate $N=4$ supermultiplets.  
The three strings also admit a soliton interpretation: one may identify the
$S$-string with the {\it elementary string} solution of \cite{Dabholkar}, the
$T$-string with the {\it dual solitonic string} solution of \cite{Khurifour} and
the $U$-string with (a limit of) the {\it stringy cosmic string} solution of
\cite{Greene}. In $D=3$ dimensions, all three strings are related by $O(8,24;Z)$
transformations.

The compactification to $N=4$, $D=4$ reveals one or two surprises: although the
$S$-string supergravity action has an off-shell $O(6,22;Z)$ which continues to
contain $SL(2,Z)_T \times SL(2,Z)_U$, the $T$-string action has only an 
off-shell
$SL(2,Z)_U \times O(3,19;Z)$ which does not contain $SL(2,Z)_S$.
Similarly, the $U$-string action has only an  $SL(2,Z)_T \times
O(3,19;Z)$ which does not contain  $SL(2,Z)_S$.  In short, none of
the actions is $SL(2,Z)_S$ invariant! This lack of  off-shell
$SL(2,Z)_S$ in the Type $II$ actions can be traced to the presence
of the extra $24$ gauge fields which arise from the Ramond-Ramond (R-R) 
sector of Type
$II$ strings: $S$-duality in the heterotic picture acts  as an on-shell
electric/magnetic transformation on all $28$ gauge fields and continues
to be an on-shell transformation on the $24$ which remain unchanged
under the string/string/string triality.  At first sight, this seems 
disastrous for
deriving the strong/weak coupling duality of the heterotic string from target
space duality of the Type $II$ string.  The whole point was to explain a {\it
non-perturbative} symmetry of one string as a {\it perturbative}
symmetry of another \cite{Duffstrong}.  Fortunately, all is not lost:
although $SL(2,Z)_S$ is not an off-shell symmetry of the Type $II$
supergravity
actions, it is still a symmetry of the Type $II$ string
theories.  To see this we first note that $D=6$ general covariance is a
perturbative symmetry of the Type $IIB$ string and therefore that the
$D=4$ Type $IIB$ strings must have a perturbative $SL(2,Z)$ acting
on the complex structure of the compactifying torus.  Secondly we note
that for both Type $IIB$ theories, $B_{TUS}$ and $B_{UTS}$, $S$ is the
complex structure field.  Thus the $T$ string has $SL(2,Z)_U 
\times SL(2,Z)_S$ and the $U$ string has 
$SL(2,Z)_S \times SL(2,Z)_T$ as 
required.  In this sense, four-dimensional string/string/string triality fills
a gap left by six-dimensional string/string duality: although duality
satisfactorily 
explains the strong/weak coupling duality of the $D=4$ Type $IIA$
string in terms of the target space duality of the heterotic string,
the converse requires the Type $IIB$ ingredient. 

Note that all of the three $SL(2,Z)_{(S,T,U)}$ take Neveu-Schwarz (NS)-NS states
into NS-NS states and that none can be identified with the conjectured 
\cite{Callan2,Hulltownsend,Wittenvarious} non-perturbative $SL(2,Z)_X$ 
of section
(\ref{Sdual}), where $X$ is the complex scalar of the Type $IIB$ theory in 
$D=10$,
which transforms NS-NS into R-R.  However, this $SL(2,Z)_X$
is a subgroup of $O(6,22;Z )$.  Since this is a perturbative target space
symmetry of the heterotic string, the conjecture follows automatically
from the $D=4$ string/string/string triality hypothesis. Thus 
we can say that  evidence for this triality is evidence
not only for the electric/magnetic duality of all three $D=4$ strings
but also for the $SL(2,Z)_X$ of the $D=10$ Type $IIB$ string and
hence for {\it all} the conjectured non-perturbative symmetries of
string theory.

We should emphasize, of course, that string/string duality and 
string/string/string
triality are themselves still only conjectures, albeit very plausible ones, 
so checks on
$S$-duality in string theory are still useful.  Trying to find all the 
$S$-dual magnetic
partners of the elementary string states is not an easy task, however, and 
seems to require
a better understanding of the role of $K3$ \cite{Gauntlett1,Porrati}.

\section{More dual string pairs}
\la{pairs}

The next conjecture to
attract attention was the one put forward by Ceresole et al \cite{Ceresole},
Kachru and Vafa \cite{Kachru} and Ferrara et al \cite{Ferrara} that the $D=10$
heterotic string compactified to $D=4$ on $K3 \times T^2$ is dual to the $D=10$
Type $IIA$ string compactified on a Calabi-Yau manifold. This yields $D=4,N=2$
dual pairs with similar symmetry enhancement properties. Taking the global limit
of this theory, we find $N=2$ Yang-Mills theories which also have the $SL(2,Z)$
and hence vanishing $\beta$ function.

Prior to this recent surge of interest in a duality between heterotic and Type
$IIA$ strings  \cite{Hulltownsend,Wittenvarious,Senssd,Harvey,Rahmfeld3}, 
however, it was
conjectured, on the basis of $D=10$  heterotic string/fivebrane duality
\cite{Duffsuper,Strominger} discussed in section (\ref{web}) below, that in 
$D\leq6$
dimensions there might exist a duality between one heterotic string and another
\cite{Luloop,Khurifour,Lublack,Minasian1,Duffclassical,Khuristring,
Duffstrong}.  
In particular, if one could find a compactification to $D=6$ for which the
heterotic string was dual to itself, then this would automatically guarantee:
(1) the quantum consistency of the dual string
\cite{Minasian1,Duffstrong}, (2) an exact symmetry between the spacetime  and
worldsheet loop expansions
\cite{Khurifour,Duffclassical,Khuristring,Duffstrong}, (3) that on further $T^2$
compactification the resulting $D=4,N=2$ theory would exhibit an exact
symmetry under interchange of the dilaton $S$ and the complex Kahler form $T$,
 and
hence (4) enhanced (non-perturbative) gauge symmetries for special values of
$S$ in
addition to enhanced (perturbative) gauge symmetries for special values of $T$
\cite{Duffstrong} as described in section (\ref{Symmetry}); a phenomenon that
does not occur in the $N=4$ theories.  

The comparative lack of interest in heterotic/heterotic duality is presumably
due  to the
lack of a convincing compactification.  The example of \cite{Minasian1}, 
where the $D=10$
$SO(32)$ heterotic string compactified to $D=6$ on $K3$ was conjectured 
to be dual to the
$D=10$ heterotic fivebrane wrapped around $K3$, seemed like a 
possible candidate, but
encountered problems with the wrong sign for some of the gauge 
kinetic terms.  In
\cite{Wittenevidence} this idea is re-examined but instead one considers the 
$E_8 \times
E_8$ string with symmetric embedding of the anomaly in the two $E_8$'s.  
It is conjectured
that this heterotic string is indeed dual to itself and that the resulting 
$D=4,N=2$
theory has the above $S-T$ interchange symmetry. This self-duality of the 
heterotic string
in $D=6$ does not rule out the possibility that in $D=4$ it is also dual to 
a Type $II$
string. In fact, as discussed in \cite{Kachru}, when the gauge group is 
completely
Higgsed, an obvious candidate is provided by the Type $IIA$ string 
compactified on a
Calabi-Yau manifold with hodge numbers $h_{11}=3$ and $h_{21}=243$, 
since this has the
same massless field content $(M,N)=(244,4)$ where $M=h_{21}+1$ 
counts the number of
hypermultiplets and $N=h_{11}+1$ counts the vectors 
(including the graviphoton).  Such a
manifold does indeed exist and is given by the degree $24$ hypersurface in
$WP^4{}_{1,1,2,8,12}$ which has recently been studied in \cite{Hosono,Klemm}. 

More recently, attention has turned to dual $N=2$ and $N=1$ dual string pairs
in $D=4$ whose global limit would correspond to the {\it effective duality}
mentioned in the Introduction
\cite{Billo,Ferrara,Kachru,Vafawitten3,Lowe,Kaplunovsky,Senvafa,Cardoso,%
Antoniadis,Klemm}. 
We will not go into the details here but simply note that although the 
field theory
duality may be only effective, the underlying string/string duality is 
exact. Interestingly
enough, all these constructions seem to involve the ubiquitous $K3$ in 
one way or another.

\section{Web of interconnections}  
\la{web}

The 1984 superstring revolution \cite{Green} answered many puzzles about the 
existence 
of a consistent perturbatively finite quantum theory of gravity and its 
unification with
the other forces.  However, it also raised some new and important questions:

i) The maximum spacetime dimension permitting a classically consistent 
supersymmetric 
field theory is $D=11$ \cite{Nahm,Julia}. Yet superstrings demand $D\leq 10$. If
superstrings  really are the theory of everything, does that mean that all 
the previous
work on $D=11$ Kaluza-Klein supergravity \cite{Pope} (including its 
compactification
on K3 \cite{Nilsson2,Nilsson1}) is of no consequence? 

ii) Ideally, one would like the ultimate theory to be unique, yet already in 
$D=10$
there were five consistent string theories: Type $IIA$, Type $IIB$, Heterotic 
$E_8 \times
E_8$, Heterotic $SO(32)$ and Type $1$ $SO(32)$. Although Heterotic 
$E_8 \times E_8$
compactified to $D=4$ on a Calabi-Yau manifold seemed most promising 
phenomenologically,
the vacuum degeneracy problem then raised its ugly head: there were 
literally billions
of such manifolds each yielding different gauge groups, different fermion
representations and different numbers of families.  How do we choose the 
right one?

iii) The $D=11$ question became more acute in 1987 when the
$D=11$ {\it supermembrane} was discovered \cite{Bergshoeff2,Bergshoeff-annp} 
and when it 
was pointed out \cite{Howe} that the $(d=2,D=10)$ Green-Schwarz action of 
the Type $IIA$
superstring follows by simultaneous worldvolume/spacetime dimensional 
reduction of the
$(d=3,D=11)$ Green-Schwarz action of the supermembrane. Could we really 
afford to ignore
supermembranes and other higher dimensional extended objects? For example, 
the $d=6$ 
worldvolume of the $D=10$ {\it fivebrane} couples to a rank six 
antisymmetric tensor
potential ${\tilde B}_{MNPQRS}$ just as the $d=2$ worldsheet of the 
string couples to
the rank two potential $B_{MN}$.  Since the $H_3=dB_2$ version of supergravity
corresponds to the field theory limit of a superstring, it was conjectured
\cite{Duffsuper} that there exists a {\it fivebrane} which may be regarded 
as fundamental
in its own right and whose field theory limit is the dual 
${\tilde H_7}=d{\tilde B}_6$
version of supergravity.  Further evidence was provided by the discovery 
of fivebrane
soliton solutions of string theory \cite{Strominger,Luelementary,Callan1,Callan2} and
evidence of a strong/weak coupling duality between the string and the 
fivebrane. Indeed
it was this {\it $D=10$ string/fivebrane duality conjecture} that 
inspired the {\it
$D=6$ string/string duality conjecture} of section (\ref{string}). 

Recent events have dramatically thrust these issues back into the limelight:

i)  In a startling paper, Witten \cite{Wittenvarious} has shown that $D=11$ 
supergravity
emerges as the strong coupling limit of the Type $IIA$ superstring and has 
conjectured a
whole web of interconnections between Type $IIA$ and the remaining four string
theories.  For example, the strong coupling limit of the $SO(32)$ heterotic 
string is
the $SO(32)$ Type $I$ string.  Further evidence for this conjecture came 
from the
discovery \cite{Dabholkar1,Hull2} that one is the soliton of the other. The 
strong
coupling limit of the $D=7$ heterotic string is $D=11$ supergravity on $K3$,
in accordance with the ``coincidence'' of the moduli in (\ref{moduli}).  
The strong
coupling limit of the $D=6$ heterotic string is the Type $IIA$ string on 
$K3$.  This
is subsumed by the string/string duality example of section 
(\ref{concrete}).  The strong
coupling limit of the $D=5$ heterotic string is the Type $IIB$ string on 
$K3$. Following
Witten's  paper \cite{Wittenvarious} it was furthermore proposed \cite{Bars}
that the combination of perturbative and non-perturbative states of the 
$D=10$ Type
$IIA$ string could be assembled into $D=11$ supermultiplets.  

ii)  Townsend \cite{Townsendeleven} has suggested that the $D=10$ Type $IIA$
superstring should be identified with the $D=11$ supermembrane compactified 
on $S^1$, with
the charged extreme black holes of the former interpreted as the Kaluza-Klein 
modes of the
latter.

iii)  Hull and Townsend showed that the (extreme electric and magnetic black 
hole
\cite{Khurinew,Rahmfeld}) Bogomol'nyi spectrum necessary for the $E_7$ 
$U$-duality of the
$D=10$ Type $IIA$ string compactified to $D=4$ on $T^6$ can be given an 
explanation in terms
of the wrapping of either the elementary $D=11$ supermembrane solution 
\cite{Stelle} or the
$D=11$ solitonic superfivebrane solution \cite{Guven} around the extra 
dimensions
\cite{Hulltownsend}.

iv)  Strominger \cite{Stromingermassless} has shown that the R-R black holes of
the Type $II$ strings can become massless and in doing so resolve the
so-called conifold singularities in the moduli space of Calabi-Yau vacua. 
In the Type IIA
theory, these black holes are nothing but membranes  which wrap around
two-surfaces of the Calabi-Yau manifold.  In the Type $IIB$ theory, they are 
threebranes
which wrap around three-surfaces. Moreover, Greene, Morrison and Strominger
\cite{Greeneblack} then showed that such black-hole condensation signals a 
smooth
transition to a new Calabi-Yau space with different topology.  Thus string 
theory unifies
the moduli space of many or possibly all Calabi-Yau vacua!

v)  The conjectured equivalence of the $D=10$ heterotic string compactified
on $T^4$ and the $D=10$ Type $IIA$ string compactified on $K3$
\cite{Hulltownsend,Wittenvarious}, combined with the above conjectures 
implies that the
$d=2$ worldsheet action of the $D=6$ ($D=7$) heterotic string may be obtained 
by $K3$
compactification of the $d=6$ worldvolume action of the $D=10$ Type $IIA$ 
fivebrane
($D=11$ fivebrane) \cite{Townsendseven,Harvey}.  

vi)  Putting all this together, one may thus conjecture that 
membrane/fivebrane duality in
$D=11$ implies Type $IIA$ string/Type $IIA$ fivebrane duality in $D=10$, 
which in turn
implies Type $IIA$ string/heterotic string duality in $D=6$. To test 
this conjecture, Duff,
Liu and Minasian \cite{Minasian2} correctly reproduced the 
corrections to the $3$-form field
equations of the $D=10$ Type $IIA$ string (a mixture of tree-level and 
one-loop effects
\cite{Vafawitten2}) starting from the Chern-Simons corrections to 
the $7$-form Bianchi
identities of the $D=11$ fivebrane (a purely tree-level effect). $K3$ 
compactification of
the latter then yields the familiar gauge and Lorentz Chern-Simons 
corrections to $3$-form
Bianchi identities of the heterotic string.

Further evidence for the importance of $D=11$ dimensions, supermembranes and
superfivebranes in superstring theory continues to appear daily on the 
internet: Schwarz
\cite{Schwarzmultiplet}, and independently Aspinwall \cite{Aspinwall4}, 
have identified
the $SL(2,Z)_X$ of the Type $IIB$ string with the modular group of the 
torus appearing in
the compactification of $D=11$ supergravity down to $D=9$.  Cadavid 
et al \cite{Cadavid},
Papadopoulos and Townsend \cite{Papadopoulos},  Schwarz and Sen 
\cite{Schwarz3}, Harvey,
Lowe and Strominger \cite{Lowe}, Chaudhuri and Lowe \cite{Chaudhuri}, Acharya
\cite{Acharya} and Aspinwall \cite{Aspinwall4} have noted that $N=1,D=4$ 
heterotic strings
can be dual to $D=11$ supergravity compactified on seven-dimensional spaces 
of $G_2$
holonomy which also yield $N=1$ in $D=4$ \cite{Pope2,Pope}.  Becker, Becker 
and Strominger
\cite{Becker} have shown that membranes and fivebranes of the Type $IIA$ 
theory, obtained
from the $D=11$ supermembrane by compactification on $S^1$, 
yield $e^{-1/g_s}$ effects,
where $g_s$ is the string coupling. Polchinski \cite{Polchinski} has 
proposed that the
Type $II$ $p$-branes carrying Ramond-Ramond charges can be given an exact 
conformal
field theory description via open strings with Dirichlet boundary 
conditions, thus
heralding the era of {\it D-branes}.  
    
It has even been shown by Horava and Witten\cite {Horava} that the 
$E_8 \times E_8$ 
heterotic string in $D=10$ may be obtained by compactifying the $D=11$ 
theory on $S^1/Z_2$
just as the Type $IIA$ string may be obtained from $S^1$.  As shown by 
Duff, Minasian
and Witten \cite{Wittenevidence}, the $D=6$ heterotic/heterotic duality 
discussed in the previous
section may then be deduced by looking in two differnt ways 
at the $D=11$
theory compactified on $K3 \times S^1/Z_2$, just as heterotic/Type $II$ 
duality may be
deduced by looking in two different ways at the $D=11$ theory compactified 
on $K3 \times
S^1$ \cite{Liu}. 

The rehabilitation of $D=11$ and the recognition of the importance of 
supermembranes
should come as no surprise to those who believe in supersymmetry: what 
is not forbidden
must be allowed. The picture that seems to be emerging from all this,
however, is that the underlying theory is very different from the 
traditional theory of
superstrings.  It is as though the theory has some enormous moduli space: 
in one corner of
moduli space it looks like a  Type $I$ $SO(32)$ string, in another corner 
like an $E_8
\times E_8$ heterotic fivebrane, in another like a $D=11$ supermembrane and 
so on. In his
book {\it Infinite in All Directions}, Freeman Dyson \cite{Dyson} 
divides theoretical
physicists into {\it unifiers} and {\it diversifiers}.  The current 
developments in duality
might be described as {\it unification via diversification}!

\section{Acknowledgements}

I have enjoyed useful conversations with the participants of the Aspen 
workshop on
duality, July 1995, and would like to thank the organizers and the Aspen 
Center for
their hospitality. Thanks are also due Ramzi Khuri, Jim Liu, Jian Xin Lu, 
Ruben Minasian,
Joachim Rahmfeld and Edward Witten.




\end{document}